\documentclass[11pt]{JHEP3}
\usepackage{graphicx,amsfonts}
\usepackage{amsmath}
\usepackage[cmtip,arrow]{xy}
\usepackage{pb-diagram,pb-xy}
\newcommand{\bea}{\begin{eqnarray}}
\newcommand{\eea}{\end{eqnarray}}

\def\com#1#2{\Big[#1,#2\Big ]}

\def\be{\begin{equation}}
\def\ee{\end{equation}}

\def\fr{\frac}
\def\a{\alpha}

\def\e{\epsilon}

\def\l{\lambda}
\def\m{\mu}
\def\n{\nu}
\def\n{\nu}

\def\D{\Delta}

\def\nn{\noindent}
\def\no{\nonumber}

\def\uq{U_q(su(2))}
\def\half{\frac{1}{2}}
\newcommand{\Hil}{\mathcal{H}}

  \def\cF{{\cal F}}

 \def\cN{{\cal N}} 
  
\def\cS{{\cal S}}

\title{Dirac operator on the q-deformed
  Fuzzy sphere and Its spectrum }
\author{E. Harikumar, Amilcar R. Queiroz and  P. Teotonio-Sobrinho\\
Instituto de F\'{\i}sica, Universidade de S\~{a}o Paulo,\\ Caixa Postal
66318, 05315-970, S\~ao Paulo, SP, Brazil\\
\email{hari@fma.if.usp.br}, \email{amilcar@fma.if.usp.br}, \email{teotonio@fma.if.usp.br}}
\abstract{ The $q$-deformed fuzzy sphere $S_{qF}^2(N)$ is the algebra
  of $(N+1)\times(N+1)$ dim. matrices, covariant with respect to the
  adjoint action of $\uq$ and in the limit $q\to 1$, it reduces to
  the fuzzy sphere $S_{F}^2(N)$. We construct the Dirac operator on the
  q-deformed fuzzy sphere-$S_{qF}^{2}(N)$ using the spinor modules of
  $\uq$. We explicitly obtain the zero modes and also calculate the
  spectrum for this Dirac operator. Using this Dirac operator, we
  construct the $\uq$ invariant action for the spinor fields on
  $S_{qF}^{2}(N)$ which are regularised and have only finite modes. We
  analyse the spectrum for both $q$ being root of unity and real,
  showing interesting features like its novel degeneracy. We also
  study various limits of the parameter space (q, N) and recover the known spectrum
  in both fuzzy and commutative sphere.}
\begin{document}
\section{Introduction}

In the fuzzy physics programme one of the key objects of interest is the
Dirac operator. It is essential for the construction of spinorial
actions on fuzzy spaces, giving, thus, kinematics of a fuzzy field.
Such field theoretic models on compact fuzzy manifolds are of interest
as they provide a regularised theory where the fields have only finite
number of modes. Also the fuzzy space being a noncommutative space,
Dirac operator is a fundamental object for the study of the
noncommutative geometry since it is one of the
ingredients required for the construction of the spectral triple
\cite{connes}. Thus apart from the field theoretic interest,
construction and study of the Dirac operator is of intrinsic interest
for the construction of differential calculus on the space being
considered.

Fuzzy spaces can be seen as highly symmetrical lattices. For this
reason we require Dirac operator to satisfy the underlying symmetry of
this space so that the field theory models constructed using it will
naturally incorporate the symmetry of the underlying space. The
prototype example is Dirac operator on the Fuzzy Sphere ($S_{F}^2$) \cite{gp,gkp,
  wattas, wattas1, bal} which is invariant under the $SU(2)$
rotations. In \cite{gp}, the Dirac operator on fuzzy sphere was
constructed and its spectrum and zero modes were obtained. This
construction was based on the generalisation of the notions of spinor
bundles to their fuzzy analogues. This was then generalised in
\cite{gkp} to include topologically non-trivial spinor modules using the
supersymmetric extension of the fuzzy sphere.  In this construction,
not only interacting fermions but also non-trivial winding number
configurations were obtained even at the kinematical level. In
\cite{wattas}, using only the bosonic algebra of fuzzy sphere, Dirac
operator and its spectrum were obtained. Here the fuzzy sphere algebra
was enlarged by including the derivations on fuzzy sphere in order to
obtain the chirality operators and Dirac operator having nice
commutative limits. Later, exploiting the ambiguity in the operator
ordering in this noncommutative generalisation, an alternative
chirality operator and the corresponding Dirac operator were obtained
in \cite{wattas1} by the same authors and their construction yields a
Dirac operator different from that obtained in \cite{gkp}. In
\cite{trg} it has been shown that the spinor theory on fuzzy sphere is
free of fermion doubling problem and further studies to construct the
topological solutions on fuzzy sphere were carried out in
\cite{balsachin}.

In \cite{hs,abiy} fuzzy generalisation of $CP^2$-space was investigated
and it was shown that the scalar theory on fuzzy $CP^2$ is free from
UV divergences and the Dirac operator on this space have many
interesting features. The investigations to understand the notions of
symmetries of noncommutative spaces based on Hopf algebras was
attempted in \cite{ps}.  Studies to construct the noncommutative
generalisations of the spaces endowed with symmetries under the action
of quantum groups have also been attempted
recently \cite{jv,apas,dlssv}. $\uq$ has also been used in the
construction of sigma model on fuzzy sphere \cite{trg1}.  A Dirac
operator on the $\uq$ group manifold was obtained and was shown to
have same spectrum as that of the round Dirac operator on a
commutative $3-$sphere. In \cite{apas} Dirac operator and chirality
operator on noncommutative space having $\uq$ as the symmetry group
were constructed. It has been argued that the Dirac operator is {\it
  covariant} and in the commutative limit where the underlying space
is Podles sphere $S_{q}^2$, the full rotational invariance of the
Dirac operator is recovered. It was further shown that the Dirac
operator reduces to that obtained in \cite{wattas,wattas1}.

Considering novel symmetries such as those related with quantum groups
is a natural follow up in the fuzzy physics programme. The quantum
groups have been already appeared in several other physical models
such as in Wess-Zumino field theories, string theory and also in knot
theory and noncommutative theory \cite{bider}. Thus the investigations
to see the possibility of constructing field theory models on
noncommutative space having invariance under the action of quantum
groups are of interest \cite{hgetal}. For the fuzzy physics programme,
where one is interested to obtain the regularised field theory with
finite number of degrees of freedom, quantum groups, specially the
$\uq$ is interesting for the introduction of one more parameter,
namely $q$ in the theory.  This allows one to study different ways of
recovering the commutative theory by different limiting procedures.

In this paper we construct a Dirac operator on q-deformed fuzzy
sphere-$S_{qF}^2$ described by an algebra ${\cal A}(N,q)$. Here $q$ is
the deformation parameter and in the limit $q\to 1$, we recover the
fuzzy sphere given by the $(N+1)\times(N+1)$ matrix algebra ${\cal
  M}(N+1)$, whose dimension is fixed by the fuzzy parameter $N$.  Here
we first construct the ${\cal A}(N,q)$ spinor-bi-modules by
generalising the construction of spinor modules to $q$-deformed case.
The spinor fields are constructed so as to have a natural
decomposition of the spinor modules into a direct sum of two
submodules which are labelled by plus and minus chirality
respectively. We also obtain a chirality operator which anti-commutes
with the Dirac operator. This is followed by the construction of a
pair of operators $K_\pm$ which maps $\pm$ chiral subspace to $\mp$ ones.
Using these $K_\pm$, a Dirac operator is constructed.  This Dirac operator maps
spinor module to itself by construction and also anti-commutes with the
chirality operator. We also require the q-deformed, fuzzy Dirac
operator and its spectrum to reduce to the known results in various
limits. These different limits can be expressed as
\bea
&S_{qF}^2\rightarrow S_{F}^2\rightarrow S^2, ~~~~(q\to 1~~{\rm
  followed~~by}~~N\to\infty)\no&\\
&S_{qF}^2\rightarrow S_{q}^2\rightarrow S^2, ~~~~(N\to\infty~~{\rm
  followed~~by}~~q\to 1)\no\\
&S_{qF}^2\rightarrow S^2, ~~~~~(N\to\infty,~~~~q\to 1 ~{\rm~~simultaneously})\no
&
\eea
where $S_{q}^2$ is the q-deformed sphere and $S_{F}^2$ is the fuzzy
sphere.  We show that the spinor modules naturally splits into the
direct sum of irreducible representations (IRR) of half-integer spins
and this allows one to obtain the eigenvectors and spectrum of Dirac
operator. We also obtain the zero modes explicitly. We present a
detailed analysis of the spectrum of the Dirac operator showing its
interesting, novel behaviour for various ranges of parameters involved
( like $q$ and fuzzy cut-off). These novel features are double
degeneracy of the spectrum for $q$ being root of unity, level crossing
of the spectrum with $q$ and cut-offs on allowed values of topological
index which are $q$ dependent. We have also shown that the spectrum in
commutative and fuzzy cases are recovered in the appropriate limits.

The present paper is organised as follows. In the next section we
present some of the essential results regarding $\uq$ as we are using
these in deriving our main result.  In section 3 we briefly revise
the construction of fuzzy spinor module and Dirac operator by Grosse
et al \cite{gkp}. In section 4 we present the main result of the
paper, viz; construction of Dirac operator and Chirality operator on
q-deformed fuzzy sphere, ${\cal A}(N,q)$. We also obtain the spectrum
of the Dirac operator and show its zero modes explicitly.  In section
5, we analyse the spectrum of deformed Dirac operator on q-deformed
fuzzy sphere showing its
interesting features and also study its various limits, recovering the
fuzzy, deformed and commutative results respectively. Our concluding
remarks are given in section 6.

\section{$\uq$, q-oscillators and Schwinger Realisation}

In this section we present the Hopf algebra $\uq$ and recall its representation theory\cite{bider}.
$\uq$ is the algebra generated by three operators, $J_{\pm}, J_3$ satisfying the relations
\bea
\com{J_+}{J_-}&=&\frac{K-K^{-1}}{q^{\frac{1}{2}}-q^{-\frac{1}{2}}},\no\\
\com{J_3}{J_{\pm}}&=&\pm J_{\pm}
\label{eq:uqdef}
\eea
where $K=q^{J_3}$. The Casimir of the algebra is given by
\begin{equation}
  \label{eq:casimir1}
  C=J_-J_++\frac{Kq^{\frac{1}{2}}+K^{-1}q^{-\frac{1}{2}}}{(q^{\frac{1}{2}}-q^{-
\frac{1}{2}})^2}.
\end{equation}
The main difference of this algebra with respect to the universal
enveloping algebra $U(su(2))$ is
the co-product. Since we will make use of tensor representation, the
co-product, antipode and co-unit play important roles. The co-product,
antipode and co-unit for $\uq$ are defined to be
\begin{eqnarray}
&\Delta (J_{\pm})=J_{\pm}\otimes K^{\frac{1}{2}}+K^{-\frac{1}{2}}\otimes J_{\pm}&\label{coproduct1} \\
&\Delta (J_3)=I\otimes J_3 + J_3 \otimes I &\label{coproduct2}\\
&S(J_{\pm})=-q^{\pm\half}J_\pm,
S(J_3)=-J_3&\\
&\e(I)=1,~~\e(J_\pm)=0,~~\e(J_3)=0\label{counit}&
\end{eqnarray}
The Schwinger realisation of $\uq$  algebra in terms of a
pair of  $q$-deformed boson operators $A_\a, A_{\a}^\dagger, \a=1, 2$
can be expressed as
\begin{eqnarray}
  \label{eq:op-definition}
  J_+ &=& A_{1}^\dagger A_2\\
  J_- &=&  A_1 A_{2}^\dagger\\
  J_3 &=& \half ( N_1 - N_2),
\end{eqnarray}
satisfying $\uq$ relations given in Eqn. (\ref{eq:uqdef}). These
q-annihilation and q-creation operators used to define the generators
of $\uq$ satisfy the following algebra ( known as q-Heisenberg
algebra)
\begin{eqnarray}
  \label{eq:qHeisalg}
  A_\a A_{\a}^{\dagger}-q^{\half}A_{\a}^{\dagger}A_{\a}&=&q^{-{\fr{N_\a}{2}}} \\
  \lbrack N_\a,A_{\a}^{\dagger}\rbrack&=&A_{\a}^{\dagger} \\
  \lbrack N_\a,A_\a\rbrack&=&-A_\a.
\end{eqnarray}
 
Depending on the values of $q$, the representations of $\uq$ can be
broadly classified into two main categories. For $q$ taking real
values, the representations are simple deformation of those of
$U(su(2))$ and they are in one-to-one correspondence with each other.
For $q$ being root of unity, there are two types of representations.
One (which we refer as classical) is similar to that of $U(su(2))$,
but finite dimensional (whose dimension is fixed by the $q$). The other
is a cyclical representation which is also finite dimensional, but not
having any relation to that of $U(su(2))$.

In the present work we are interested only in cases where $q$ is root
of unity or real, but with classical representation. In this
settings we will study certain limits where we recover the usual
situation of $U(su(2))$.

For both types of representations we are interested, we have,
\begin{eqnarray}
  \label{eq:irrep}
  J_{\pm}|l,m\rangle&=&\sqrt{[l\pm m +1]_q[l\mp m]_q}~|l,m\pm 1\rangle \\
  J_3|l,m\rangle&=& m |l,m\rangle. 
\end{eqnarray}
Here we have used the q-number defined as 
\be
[n]_q\equiv [n]= \fr{{q^{\fr{n}{2}}} -{q^{-\fr{n}{2}}}}{q^{\half} -q^{-\half}}
\label{qno}
\ee
\section{Dirac operator on fuzzy sphere}\label{fuzzy}

Here we briefly summarise the essential results regarding the
construction of spinor modules and $su(2)$ invariant Dirac operator
\cite{gkp} on the fuzzy sphere $S_{F}^2$. Apart from fixing our
notations, it also serves to compare and contrast the construction of
the Dirac operator and its spectrum on fuzzy sphere-$S_{F}^2$ and that
on q-deformed fuzzy sphere-$S_{qF}^2$.

The spinor fields on $S_{F}^2$ is defined as
\be
\Psi = f(a_{\a}^\dagger, a_\a)b +g(a_{\a}^\dagger, a_\a)b^\dagger.
\label{spin}
\ee
Here $b$ and $b^\dagger$ are fermionic annihilation and creation
operators and $f$ and $g$ are constructed using a pair of
bosonic annihilation and creation operators, $a_\a, a_{\a}^\dagger$
alone. Thus both $f$ and $g$ have odd Grassmann parity. The above
bosonic and fermionic operators obey $a_{1}^\dagger a_{1}
+a_{2}^\dagger a_{2}+b^\dagger b= N$ where $N$ is the fuzzy cut-off
parameter. 

The the spinor field $\Psi$ given above can be considered as the
linear combination of monomials (given below in Eqn. (\ref{monomial}))
with fixed topological index $2k=m_1 +
m_2+\m - n_1 - n_2 -\n,~\m,\n=0,1~{\rm and ~} ~ k\in
\half \mathbb{Z}$. Here
$\m,\n$ represent the number of fermionic creation and annihilation
operators appearing in Eqn.(\ref{spin}) and $m_1, m_2, n_1, n_2$ are non-negative integers.
The space of these spinors is
denoted as ${\cS}_k, k\in \half \mathbb{Z}$ and the $su(2)$ algebra has a
natural action on these modules. Since $\m,\n=0,1$ and $\m+\n=1$, we
notice that the spinor module naturally splits into two parts
\be
{\cal S}_k={\cal S}_{k+\half} b + {\cal S}_{k-\half}b^\dagger
\ee
and $f$ and $g$ belonging to ${\cS}_{k\pm\half}$
can be generically expressed as
\be
\phi=\sum C_{m_1 m_2 n_1 n_2}
a_{1}^{\dagger {m_1}}a_{2}^{\dagger {m_2}}a_{1}^{n_1}
a_{2}^{n_2}.
\label{mono}
\ee
respectively. Here $m_1, m_2, n_1, n_2$ are non-negative integers.

Note that these bi-modules ${\cS}_k\equiv{\cS}_{M,N}$ ($su(2)$ acts on
them by left-right actions) are generated by
the monomials of the form
\be
a_{1}^{\dagger {m_1}}a_{2}^{\dagger {m_2}}a_{1}^{n_1}
a_{2}^{n_2}b^{\dagger \m} b^{\n}
\label{monomial}
\ee
with $m_1+m_2+\m\le M, n_1+n_2+\n
\le N, M-N=2k$ apart from the condition $(m_1+m_2+\m)-(n_1+n_2+\n)=2k$
and they map the Fock space ${\cF}_{N}^\n$ to
${\cF}_{M}^\m$. Here, the
elements of the space ${\cF}_{N}^\n$ are given by
\be
|n_1, n_2; \n> =\fr{1}{\sqrt{n_{1}!n_{2}!}} a_{1}^{\dagger {n_1}}
a_{2}^{\dagger {n_2}}b^{\dagger \n}|0>, ~~n_1+n_2+\n=N
\label{focksp}
\ee where $|0>$ is the vacuum state, $n_1, n_2$ are non-negative
integers and $\n=0,1$. Thus we see that any operator
$\Phi\in{\cS}_{M,N}$ can be expressed as a $(2N+1)\times(2M+1)$
matrix. From the above discussion, we see that for a given value of
$k$, $f\in {\cS}_{M,N-1}$ and $g\in {\cS}_{M-1,N}$.  Also, the spinor
operators can be seen to map different Fock space as 
\bea
f(a_{\a}^\dagger, a_\a)b:{\cF}_{N}^{1}\to{\cF}_{M}^{0}\no\\
g(a_{\a}^\dagger, a_\a)b^\dagger: {\cF}_{N}^{0}\to{\cF}_{M}^{1}
\eea
where the element of ${\cF}_{N}^\n$ are defined by Eqn. (\ref{focksp}).  Thus it is
clear that $f$ and $g$ can be expanded in terms of the tensor
operators belonging to the half-integer spin representations\cite{gkp}
\bea
\fr{M}{2}\otimes\fr{N-1}{2}=|k+\half|\oplus......\oplus(J-\half)\\
{\rm ~and~}~~~~ \fr{M-1}{2}\otimes
\fr{N}{2}=|k-\half|\oplus......\oplus(J-\half) \eea respectively where
$J$ is related to $M$ and $N$ by $M+N=2J$.

From the above discussion, we see that the tensor operators belonging to the
bi-modules ${\cS}_{k+\half}$ and ${\cS}_{k-\half}$, i.e., $f$ and $g$ can
be expressed as a linear combination of state vectors corresponding to
the IRRs of the half-integer spins ranging from $|k-\half|$ up to
$(J-\half)$ respectively. To this end, we first construct the state vectors
for these IRRs of the half-integer spins with arbitrary value of
$k$.

For a given $j$ and $k$ we verify that the lowest weight state is
\be
\Phi_{J,k,-j}^{j}= {\cN} a_{2}^{\dagger (j+k)} a_{1}^{~ (j-k)}
\ee
and the state vectors $\Phi_{J k m}^{j}$ corresponding to other values of $m$ can be obtained by
the action of $J_+$. 
We also note that $fb$ and $gb^\dagger$ are separately eigenfunctions of the
Chirality operator $\Gamma$ with eigenvalues $\pm 1$. The chirality
operator $\Gamma$ defined using the fermionic operators $b, b^\dagger$
acts on the spinors as
\bea
\Gamma \Psi &=& -[b^\dagger b, \Psi]\no\\
&=& f(a_{\a}^\dagger, a_\a)b - g(a_{\a}^\dagger, a_\a)b^\dagger.
\eea
This chirality operator allows a natural splitting of the space of
spinor fields $\Psi$ (see Eqn. {\ref{spin}}) according
to chirality. Using the state vectors $\Phi_{J k m}^{j}$, we define the
$\pm$ chiral spinors as
\bea
\Phi_{J,k,m}^{j +}&=&\Phi_{J,k+\half,m}^{j} b\\
\Phi_{J,k,m}^{j -}&=&\Phi_{J,k-\half,m}^{j} b^{\dagger}.
\eea

The Dirac operator is defined as an operator mapping the spinor module
${\cal S}_k$ to itself and having invariance under $U(su(2)$ \cite{gkp}.  It
can be written in terms of two operators $K_\pm$ and its action is
defined as
\be
D\Psi=(K_{+} g(a_{\a}^\dagger, a_\a)b)+(K_{-} f(a_{\a}^\dagger, a_\a)b^\dagger).\ee
Here, $K_\pm$ are the operators mapping the spinors from $\pm$ chiral
subspace to $\mp$ chiral subspace and their action on a generic
vector $\phi$ are given by
\bea
K_+\phi&=&b a_{2}^\dagger \phi a_{1}^{\dagger}b-ba_{1}^\dagger \phi
a_{2}^{\dagger}b\\
K_-\phi&=&b^\dagger a_{1} \phi a_{2} b^\dagger -b^\dagger a_{2} \phi a_{1} b^\dagger
\eea
We also note that the chirality operator $\Gamma$ anti-commute with
the Dirac operator. This guarantees the chiral invariance of the
spinor field action constructed using this Dirac operator.

Using the chiral spinors constructed above, we obtain the eigenvectors
of Dirac operator $ \Psi_{J,k.m}^{j\pm}= 
\fr{1}{\sqrt{2}}\left[ \Phi_{J, k, m}^{j+}\pm  \Phi_{J, k,
    m}^{j-}\right]$ with eigenvalues $\sqrt{(j+\half+k)(j+\half-k)}$.

\section{Dirac operator on q-deformed fuzzy sphere }\label{qfuzzy}

In this section we obtain the Dirac operator on the q-deformed fuzzy
sphere. This is done by first constructing the ${\cal A}(N,q)$ spinor bi-modules
which can be expressed as the direct sum of IRRs of half-integer
spins. Here we show the role of the parameter $q$ in deciding the
maximal spin along with the fuzzy cut off ${\cal N}$. We next show that the
eigenfunctions spanning these half-integer spaces, along with
fermionic creation and annihilation operators which define a chiral
operator, provide a natural splitting of the spinor bi-modules into
two submodules characterised by the eigenvalues of chirality operator.
We also obtain a pair of operators $K_\pm$ that maps from $\pm$ chiral
subspace to $\mp$ chiral subspace and show that the eigenfunctions
spanning the half-integer IRRs are also eigenfunctions of $K_\pm$.
Using these operators, we then construct the Dirac operator and its
eigenfunctions and spectrum are obtained. We also obtain the zero
modes of the Dirac operator.

\subsection{$\uq$ spinor module}

The q-deformed spinor belonging to the spinor module ${\cal
  S}_{k}\equiv {\cal S}_{MN},
2k=(m_1+m_2+\m -n_1-n_2-\n), k\in \half \mathbb{Z}$ is given as
\be
\Psi=f(A_{\a}^\dagger, A_\a)b+g(A_{\a}^\dagger A_\a)b^\dagger
\label{qspinor}
\ee
where $b, b^\dagger$ are fermionic annihilation and creation operators
and $A_\a, A_{\a}^\dagger$ are the corresponding q-deformed
bosonic operators using which $f$ and $g$ are constructed.
The fermionic and q-deformed bosonic operators are related to the
fuzzy cut-off parameter ${\cal N}$ by  the relation $[N_1]+ [N_2] +b^\dagger
b={\cal N}$, where $N_\a$ are the q-number operators.
The operators $f$ and $g$, having odd Grassman parity, can be
generically expressed as
\be
\phi=\sum C_{m_1 m_2 n_1 n_2}
A_{1}^{\dagger {m_1}}A_{2}^{\dagger {m_2}}A_{1}^{n_1}
A_{2}^{n_2}
\ee
where $m_1, m_2, n_1, n_2$ are non-negative integers. Thus the spinor
field $\psi$ in Eqn. (\ref{qspinor}) can be expressed as a linear
combination of monomials involving q-creation and q-annihilation operators
and {\it usual} fermionic creation and annihilation operators.
These monomials having fixed topological index $2k$ are generically expressed as
\be
A_{1}^{\dagger {m_1}}A_{2}^{\dagger {m_2}}A_{1}^{n_1}A_{2}^{n_2}b^{\dagger \m}b^{\n}
\ee
with $m_1+m_2+\m\le M, n_1+n_2+\n\le N, M-N=2k$ and also
$(m_1+m_2+\m)-(n_1+n_2+\n)=2k$ where $k\in \half \mathbb{Z}$. In the
above, $\m,\n$ denote the number of fermionic creation and
annihilation operators. Since $\m,\n=0,1$, and $\m+\n=1$,
spinor field naturally decompose into $\pm$ chiral 
components and the q-tensor operators $f$ and $g$ belong to the
bi-modules ${\cal S}_{k\pm\half}$ respectively (see also the discussions after
Eqn. (\ref{spin})).

It is
easy to see from the above that the bi-modules ${\cal S}_{k}\equiv {\cal S}_{MN}$ satisfy
$$S_{MN}S_{NO}=S_{MO},~~~ S_{MN}^\dagger=S_{NM}$$ as in the
former(undeformed) case. 
These bi-modules $S_{MN}$ maps the Fock spaces ${\cF}_{N}^\n\to{\cF}_{M}^\m$. 
These (finite dimensional) Fock spaces are generated by bosonic
q-creation operators and fermionic creation operator and is given by
\be
|n_1, n_2, \n>_q=\fr{1}{\sqrt{[n_1]![n_2]!}}A_{1}^{\dagger
    {n_1}}A_2^{\dagger{n_2}}b^{\dagger\n}|0>, n_1+n_2+\n=N
\ee
where $|0>$ is the vacuum state annihilated by $A_\a$ and $b$. Thus the
operators belonging to $S_{MN}$ can be expressed as
$(2N+1)\times(2M+1)$ matrices. 
The q-tensor operators $f$ and $g$ which forms the spinor in
eqn. (\ref{qspinor}) belongs to the bi-modules $S_{k\pm\half}$ and
these bi-modules can be written as the direct sum of IRRs
corresponding to half-integer spins as
\bea
\fr{M}{2}\otimes\fr{N-1}{2}&=&|k+\half|\oplus......\oplus(J-\half)~~({\rm
  for }~f)\label{f}\\
{\rm ~and~}~~~\fr{M-1}{2}\otimes\fr{N}{2}&=&|k-\half|\oplus......\oplus(J-\half)~~({\rm for}~g)\label{g}
\eea
respectively and here $J=\fr{M+N}{2}$ and $k=\fr{M-N}{2}$ as in the
undeformed case. Thus the spinor module can be expressed as a sum of
IRRs of $\uq$. But here q-Clebsh-Gordan coefficients appear in the tensoring of
eigenfunctions unlike in the undeformed case where it is governed by
the Clebsh-Gordan coefficient. Here we note that by fixing $M+N$ where
$M$ and $N$ are the upper cut-off on the total number of creation and
annihilation operators, one effectively fixes the fuzzy cut-off. Thus
the maximum allowed value of $J$ is restricted by the fuzzy cut-off
parameter ${\cal N}$.

The set of these spinor fields, ${\cal S}_{k}$ is
$\uq$-algebra bi-module and thus has a natural action of the algebra
on them. This action is defined by the co-products
given in Eqns.(\ref{coproduct1},\ref{coproduct2}).
From this co-product, we verify the action of $\uq$ generators $J_\pm$ on an
irreducible tensor operator $\Phi_{m}^j$ as
\be
\D(J_\pm)\Phi_{m}^j= (J_\pm \Phi_{m}^j-q^{-\fr{m}{2}}\Phi_{m}^j
J_\pm)q^{-\fr{J_3}{2}}=\sqrt{[j\pm m+1][j\mp m]}~\Phi_{m\pm 1}^j
\label{cop}
\ee
and using this, we obtain the lowest weight state for each of the
above half-integer spin representations. Thus we see, for each given
values of $j$ and $k$, the lowest state is 
\be
\Phi_{J,k,-j}^{j}=(A_{2}^{\dagger} q^{\fr{N_1}{4}})^{(j+k)}
(A_{1}q^{-\fr{N_2+1}{4}})^{(j-k)} .
\label{qlowest}
\ee
We can construct the remaining state
vectors for all other values of $m$ (for a fixed value if $k$) by
applying the $J_+$ on this $\Phi_{J, k,-j}^{j}$ [We can also start
from the highest weight state given by
$\Phi_{J,k,j}^{j}=(A_{1}^\dagger q^{-\fr{N_2}{4}})^{(j+k)}
(A_{2}q^{\fr{N_1+1}{4}})^{(j-k)}$ and obtain the remaining states by
applying $J_-$].

\subsection{Chiral spinors, Chirality operator and $K_\pm$}

The components of the fermionic field defined in Eqn. (\ref{qspinor}),
viz; $fb$ and $gb^\dagger$ can be expressed using
the above obtained state vectors belonging to IRRs of half-integer
spins and the fermionic creation and annihilation operators $ b^\dagger$
and $b$. We can expand these chiral
components $f b$ and $g b^\dagger$ in terms of 
\bea
\Phi_{J,k,m}^{j+}&=&\Phi_{J,k+\half,m}^{j}b, ~j=|k+\half|,....,(J-\half)\label{pchiral}\\
{\rm and}~~~~\Phi_{J,k,m}^{j-}&=&\Phi_{J,k-\half,m}^{j}b^\dagger
~ j=|k-\half|,....,(J-\half)
\label{pmchiral}
\eea
respectively. Using the chirality operator $\Gamma$ having the same
form as that in the undeformed case, we see that $f b$ and $g b^\dagger$ defined
above are eigenvectors of $\Gamma$ with eigenvalues $\pm 1$ and its
action on spinor is given by
\be
\Gamma\Psi =-[b^\dagger b, \Psi].
\ee

The Dirac operator is required to map the spinor modules ${\cal S}_{k}$ to
itself and also required to anti-commute with the above chirality
operator. Further, we require the Dirac operator to be invariant in the sense of Eqn. (\ref{inv}).
We can construct the Dirac operator satisfying the above
requirements if we can provide two operators $K_\pm$ which will swap
the elements of $\pm$ chiral subspace to that of $\mp$. That is, the
$K_\pm$ operators should be such that
\be
K_\pm \Phi_{J,k,m}^{j\mp}\to\Phi_{J,k,m}^{j\pm}
\label{swap}
\ee
and then using these operators, we can construct the Dirac operator
with the required properties. These operators, $K_\pm$ satisfying Eqn.
(\ref{swap}) are given by
\bea
K_+\Phi&=&q^{-\fr{k- m}{4}} q^{-\fr{J_z}{2}} b \left [ A_{1}^\dagger \Phi
  A_{2}^\dagger q^{\fr{k}{2}}-  A_{2}^\dagger \Phi
  A_{1}^\dagger\right]b \\
K_-\Phi&=& q^{-\fr{k+ m}{4}} b^\dagger \left [ A_{1} \Phi
  A_{2} q^{\fr{k}{2}}-  A_{2}\Phi
  A_{1}\right]b^\dagger q^{-\fr{J_z}{2}}.
\eea
Using the above and Eqn.(\ref{qlowest}), we obtain
\be
K_\pm \Phi_{J,k,m}^{j}= \sqrt{[j\pm k+1][j\mp k]}~\Phi_{j,k\pm 1, m}^{j}
\label{keigen}
\ee

\subsection{Dirac operator and its Spectrum}

Next we use these operators $K_\pm$ to construct the Dirac operator
which guarantees that the Dirac operator maps spinor module to itself.
The action of the Dirac operator is expressed as
\be
D\Psi= K_+\Phi_{J,k,m}^{j-} +K_-\Phi_{J,k,m}^{j+}.
\label{diracop}
\ee
Using the Eqns. (\ref{pchiral}, \ref{pmchiral}, \ref{keigen}) and
(\ref{diracop}), we obtain,
\be
D\Phi_{J,k.m}^{j\pm}=\sqrt{[j+\half+k][j+\half-k]}~\Phi_{J,k.m}^{j\pm}
\label{deigen}
\ee
Thus we see that the normalised eigenfunctions of the Dirac operator
can be  written as a linear combination of $\pm$ chiral states as
\be
\Psi_{J,k.m}^{j\pm}= \fr{1}{\sqrt{2}}\left[ \Phi_{J, k, m}^{j+}\pm  \Phi_{J, k, m}^{j-}\right] 
\ee
with eigenvalues $\l(j,k)=\sqrt{[j+\half+k][j+\half-k]}$. From Eqns. (\ref{f})
and (\ref{g}), we note that the allowed values of $j$ in the above
ranges between $j=|k|+\half$ to $j=J-\half$. 

We also see that the $|M-N|$ zero modes of Dirac operator are
\bea
\Psi_{+0}^{m_1 m_2} &=& A_{1}^{\dagger {m_1}}A_{2}^{\dagger {m_2}}b^\dagger\\
\Psi_{-0}^{n_1 n_2}&=& A_{1}^{n_1} A_{2}^{n_2}b .
\eea
For the zero modes $\psi_{+}$ the allowed values for the index are
$k=\half(m_1+m_2+1)>0$ and for $\Psi_{-}$ it is given by
$k=-\half(n_1+n_2+1)<0$ and in both cases $j=|k|-\half$. We also
mention that the above
spectrum satisfy the index theorem and a detailed analysis of this
issue will be reported separatly.

\subsection{Spinor and $\uq$ Invariance}
Using the co-product defined in Eqns.
(\ref{coproduct1},\ref{coproduct2},\ref{cop}) and Eqn. (\ref{keigen}), it is
easy to see that the chiral spinors defined in Eqns. (\ref{pchiral})
and (\ref{pmchiral}) obey
\be
\D{J_\pm}(K_\pm \Phi)- K_\pm(\D(J_{\pm})\Phi)=0.
\label{inv}
\ee
To give an explicit example, we consider the simplest case of
$j=\half=k$ where the states are
$\Phi_{J,\half,-\half}^{\half}=A_{2}^\dagger q^{\fr{N_1}{4}},
\Phi_{J,\half,\half}^{\half}=A_{1}^\dagger q^{\fr{-N_2}{4}}$ and 
$j=\half=-k$ for which the states are
$\Phi_{J,-\half,-\half}^{\half}=-A_1 q^{\fr{-(N_2+1)}{4}},
\Phi_{J,\half,\half}^{\half}=A_{2} q^{\fr{(N_1+1)}{4}}$. Here the
$J_\pm$ operators act (through the co-product) as
raising and lowering operators for $k=\half$ and $k=-\half$
sets separately and $K_\pm$ maps from one set to another but keeping the $m$
value unaltered. This property of $\Phi$s given in Eqn. (\ref{inv})
guarantees the invariance of the action constructed using the
spinor fields $\Psi$ as we will see below.

The spinor field $\Psi$ can be expressed as a linear combination of the
zero modes and $\pm$ chiral spinors as
\be
\Psi=\sum_{m_1,m_2} C_{m_1 m_2}^{\pm} \Psi_{\pm 0}^{m_1 m_2} +
\sum_{j=|k|+\half}^{J-\half} \left ( C_{km}^{j+}\Phi_{J, k,m}^{j+}+ C_{km}^{J-}\Phi_{J,k,m}^{j-}\right)
\ee
where the $C$s are coefficients. Using this $\Psi$ and
$\Psi^\dagger(=bg^\dagger-b^\dagger f^\dagger)$ and the Dirac
operator, the spinor action can be defined as shown below.
\subsection{ Invariant Trace and Spinor Action on $S_{qF}^2$}

The invariant action for the spinorial field on fuzzy sphere is
defined using the trace-$Tr$ which is invariant under the action of
underlying algebra-$su(2)$. In the case of the fuzzy sphere this trace
was the usual trace of matrices. In the present case where the
symmetry algebra is $\uq$, the trace defined in this way is not invariant.
Thus we need to define a Trace operation satisfying
\begin{equation}
  \label{invtrdef}
  Tr(\Delta(a)\hat{A})=\epsilon(a)Tr(\hat{A}),
\end{equation}
where $a$ are the generators of $\uq$ and $\e$ is the co-unity defined
in Eqn. (\ref{counit}). Here $\hat{A}$ is a matrix defined using the tensor 
operators of
$\uq$ as
\begin{equation}
  \label{eq:tensor-operator}
  \hat{A}=\sum_{j=0}^{N}\sum_{k=-j}^{j} A_{jk}\hat{T}_{jk}.
\end{equation}
The non-invariance of the
usual trace is because of the co-product appearing in
Eqn. (\ref{invtrdef}). Following the general prescription of
the deformation of the trace \cite{Chari} we define a new trace which
is invariant under $\uq$. For this,
instead of $\hat{A}$ belonging to the representaion space
$\Hil_l\otimes\Hil_l^*$, we consider the matrix $\hat{A}^b$ in the equivalent
representation space $\Hil_l^{**}\otimes\Hil_l^*$.  Thus define
prescription for the $q-$trace as
\begin{equation}
  \label{eq:q-trace}
  Tr_q(\hat{A})=Tr({\hat K}\hat{A})
\end{equation}
where ${\hat K}$ is the matrix representation of $K=q^{J_3}$. It can be easily verified that the above trace satisfies the
invariance condition given in Eqn. (\ref{invtrdef}). Using this
invariant trace, 
the action for spinorial fields with topological index $k$ is given
by:
\begin{equation}
  \label{eq:k-Dirac-action}
  S_{2k}=\frac{2\pi R^2}{[N+1]}Tr_q\left[{\bar\Psi} D\Psi+V({\bar\Psi}\Psi)\right],
\end{equation}
where $\frac{2\pi R^2}{[N+1]}$ is a normalization factor, with $R$ being
associated with the radius of the underlying sphere. For arbitrary
topological index we have:
\begin{equation}
  \label{eq:Dirac-equation}
  S=\sum_{2k}S_{2k}
\end{equation}
 
Thus, using the new definition of the trace, we obtain an 
$\uq$ invariant action (\ref{eq:k-Dirac-action}) for spinorial fields.

\section{Spectrum and its Limits}\label{spec}

In this section, we analyse the spectrum of the Dirac operator on the
q-deformed fuzzy sphere and study its various limits.

The spectrum of the Dirac operator given in Eqn. (\ref{deigen}) depends
on $j$ and $k$. But unlike the commutative and/or fuzzy cases, here it
depends on $q$ also. Here we analyse the response of the spectrum to
changes in these parameters. We also study various limits of this
spectrum. We consider two main classes, viz; $q$ being root of unity
and real. In both these cases, we also study the behaviour of the spectrum
for both vanishing and non-vanishing topological index. Since the
eigenvalues for both chiral spinors $\Phi_{J,k.m}^{j\pm}$ are same,
without lose of generality, we consider only the positive values of $2k$.

\subsection{The case of $q$  root of unity}

\begin{figure}[htb]
\begin{center}
\includegraphics[height=9cm,width=11cm]{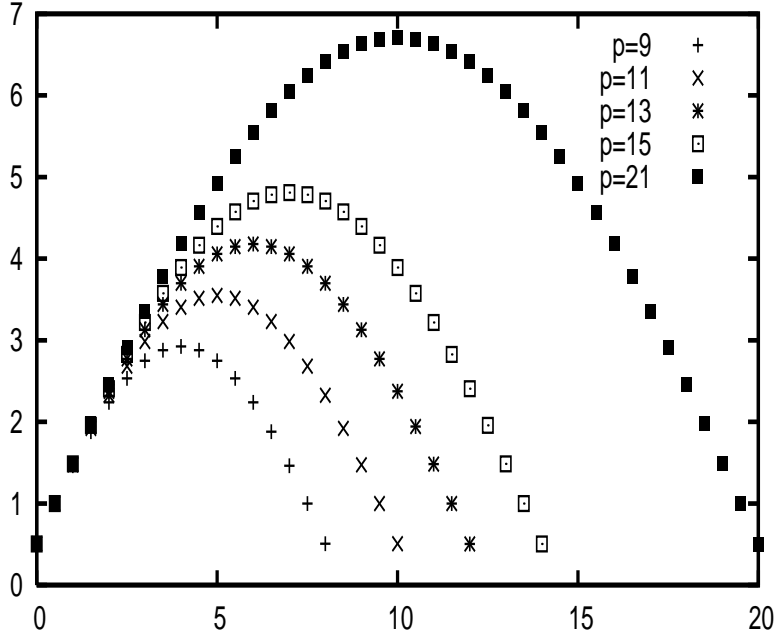}
\end{center}
\caption{Spectrum for $q=e^{\fr{2\pi i}{p}}$  and $2k=0$ as a function of spin $j$ for different values of $p$.}\label{graf1}
\end{figure}

First we consider the case were $q$ is root of unity, i.e.,
$q=e^{\fr{2\pi i}{p}}, p\in \mathbb{Z}$. Using this in Eqn. (\ref{qno}), we see
that
\begin{equation}
  \label{eq:q-number}
  [x]=\frac{sin(\frac{\pi x}{p})}{sin(\frac{\pi}{p})}.
\end{equation}
From this, it is clear that $[N+1]=0$ if $p=N+1$. This shows that the
fuzzy cut-off parameter $N$ (which fixes the matrix dimension of the
operators to be $(N+1)\times (N+1)$) has to satisfy the constraint
\begin{equation}
  \label{eq:constraint}
  N+1<p.
\end{equation}
With this condition on $q$, we first
consider the spectrum for topological index, $2k=0$ case. In Fig. 
(\ref{graf1}) we show the variation
of the spectrum for varying $j$ (for different values of $p$).
From the Fig.(\ref{graf1}) we notice that
spectrum is not monotonically increasing, unlike in the case of 
the Fuzzy Sphere$S_{F}^2$ (where $Spec(D_F,k=0)=j$).
The striking feature of the spectrum is its {\it new} degeneracy (apart
from the one with respect to the quantum number $m$) which is not shared either by fuzzy or by commutative
counterparts. This can be easily seen by considering the 
plot of spectrum in fig. (\ref{graf1}) for $p=9$, for example. In this case the
maximum value $N$ can take is 7 and we see for $j$ taking values up to 
$\fr{(N+1)}{2}$, there is no `degeneracy' in the spectrum. But as soon as $j$ is
allowed to take values above $\fr{(N+1)}{2}$, {\it new} degeneracy sets in.
Thus, in general we see that for a fixed $p$ (and $N<p-1$),
the spectrum becomes {\it doubly} degenerate once
$N>\frac{p-1}{2}$. For the example of $p=9$, we show this double
degneracy explicitly in the Table-I where the eigenvalues of
Dirac operator $\l(j,k)$ is given for different values of $j$ with $k=0$. 
\begin{table}
\begin{center}
\begin{tabular}{|c|c|}
\hline\hline
Degenerate~ levels& Value of\\
for $ p=9$&Spectrum\\
\hline
$\l(0, 0)=\l(8,0)$ &0.507713\\
\hline
$\l(\half,0)=\l(\fr{15}{2}, 0)$ &1.000000\\
\hline
$\l(1,0)=\l(7,0)$ &1.41902\\
\hline
$\l(\fr{3}{2},0)=\l(\fr{13}{2}, 0)$ &1.879385\\
\hline
$\l(2, 0)=\l(6, 0)$ &2.239764\\
\hline
$\l(\fr{5}{2}, 0)=\l({\fr{11}{2}}, 0)$ &2.532089\\
\hline
$\l(3,0)=\l(5, 0)$ & 2.747477\\
\hline
$\l(\fr{7}{2},0)=\l(\fr{9}{2}, 0)$ &2.879385\\
\hline
$\l(4, 0)$  & 2.923804\\
\hline\hline
\end{tabular}
\end{center}
\caption{Eigenvalues $\l(j,k)$ for different values of $j$ with $k=0$}
\end{table}
\begin{figure}[htb]
\begin{center}
\includegraphics[height=7cm,width=9cm]{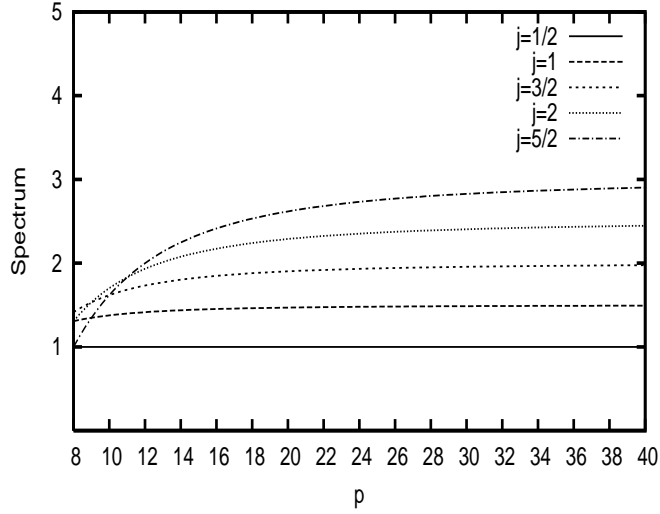}
\end{center}
\caption{Variation of the spectrum with $p$ for different values of $j$
  for $q=e^{\fr{2\pi i}{p}}$ and $2k=0$.}\label{graf2}
\end{figure}
\begin{figure}[htb]
\begin{center}
\includegraphics[height=7cm,width=9cm]{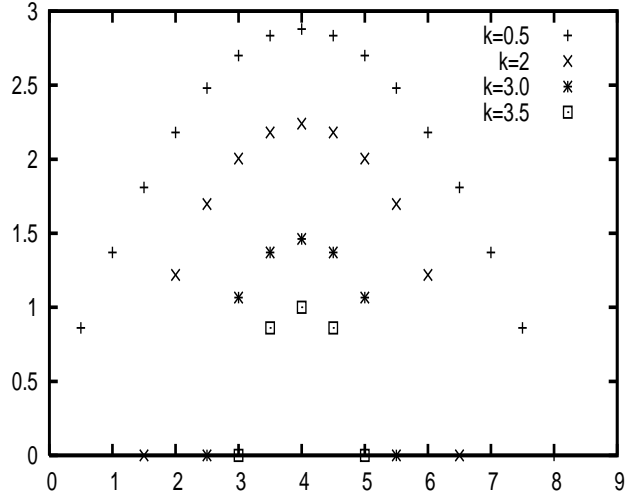}
\end{center}
\caption{Variation of the spectrum with $j$ for different values of $k$
for $q=e^{\fr{2\pi i}{p}}$ and $p=16$. }\label{graf3}
\end{figure}

In the next graph (fig.\ref{graf2}), we plot the spectrum for given
values of $j$ but as a function of $p$. For small values of $p$, we
see the spectrum has a different behaviour and it shows a growth with
increasing $p$. For small $j$ but large $p$, the spectrum behaves very
similar to that of the fuzzy case. From this graph, we also note that
the spectrum corresponding to the higher values of $j$ lie below that
of smaller $j$ when $p$ is low. As $p$ increases, there is a {\it
  level crossing} and the spectrum for higher $j$ raise above that of
smaller $j.$

Next we plot the variation spectrum with $j$, but for fixed,
non-vanishing topological indices $2k$ in (Fig.\ref{graf3}). We
notice that the spectrum is degenerate for non-vanishing values of $k$
also. More interestingly we see that the spectrum becomes zero for
certain values of $j$ for each given values of $k$. But since we are
dealing with non-zero modes only, this imply that all values of $k$
are not available for every spin $j$. The restrictions on $k$ will
become clear below where we analyse various limits of the spectrum.

First we consider the limit of $p \to \infty$ with $N$ fixed. Since in
the limit $p \to \infty$ we have $q \to 1$ and the $q$-number goes to
the usual number,i.e., $[x]\to x.$ Thus we expect to recover the
spectrum of fuzzy Dirac operator and we see below that this is true.
Considering first the case for vanishing topological index, the
eigenvalue equation for the $q$-Dirac operator
\begin{equation}
  \label{eq:qdirac-eigen-k=0}
  D^q\Phi_{J,0,m}^{q,j}=[j+\half]\Phi_{J,0,m}^{q,j},
\end{equation}
showing that for the highest value of $j$, the eigenvalue is 
$[\fr{N}{2}+\half]$.
Now, taking the limit $p\to \infty$ with $N$ fixed, it becomes
$(\fr{N}{2}+\half)$, giving the result of fuzzy case. It
is easy to see that the same happens for every value of $j<N$.

Next, for the case of non-vanishing topological index, consider the
eigenvalue equation given in Eqn. (\ref{deigen}). Since we are not
considering the zero-modes, each of the factors in the eigenvalue have
to be non-zero which lead to the inequalities
\bea
\fr{N+1}{2}-k\ne 0\label{eq:constraint3}\\
 \fr{(N+1)}{2}+k<p,\label{eq:constraint2}
\eea
Eqn. (\ref{eq:constraint3}) sets a lower cut-off for the allowed values
of $N$ for fixed $k$. For a fixed $N(=\beta)$, Eqn.
(\ref{eq:constraint2}) and Eqn. (\ref{eq:constraint}) together imply $
k< p-\fr{(\beta+1)}{2}$. Since the highest allowed value of $\beta$
satisfying Eqn. (\ref{eq:constraint}) is $\beta=p-2$, we get
\begin{equation}
 k<\fr{(p+1)}{2}.
\end{equation}
Thus we see here that $p$ acts as an upper cut-off for the topological
index $2k$. Thus by fixing $p$, we not only impose an upper cut-off for the 
fuzzy parameter $N$ but also restrict the allowed values for
topological index. This feature is particular to q-deformed fuzzy
sphere with q being root of unity.

We also note that, as in the $k=0$ case, here too, in the limit of $p \to
\infty$ with fixed $N$, the spectrum becomes that of the fuzzy Dirac
operator.

Next we consider the case where $N\to \infty$, $p \to \infty$,
$\frac{N}{p} = \alpha$ (say, with $0<\alpha< \half$). In this way, one
passes from the $q$-deformed fuzzy sphere to the commutative sphere.
Since $p \to \infty$ imply $q \to 1$ and thus in this limit, the
spectrum goes to that of the commutative sphere as the fuzzy cut off
$N$ is sent to $\infty$. This happens for both zero and non-zero
values of $2k$.

\subsection{The case of  the $q$ being real}

Here we study the case when the deformation parameter $q$ is real,
i.e., $q=e^{2p}$. In this case, from Eqn. (\ref{qno}) we see
that the q-number can be expressed as 
\begin{equation}
[x]=\fr{sinh(xp)}{sinh(p)}.
\end{equation}
Thus, we see that $p$ does not introduce any restriction on the fuzzy
cut-off N, unlike in the situation where q was root of unity. Thus in
the present case, the cut-off N can be arbitrarily large and thus we
can take the limit of $N\to\infty$ independently of the parameter $q$.
\begin{figure}[htb]
\begin{center}
\includegraphics[height=7cm,width=9cm]{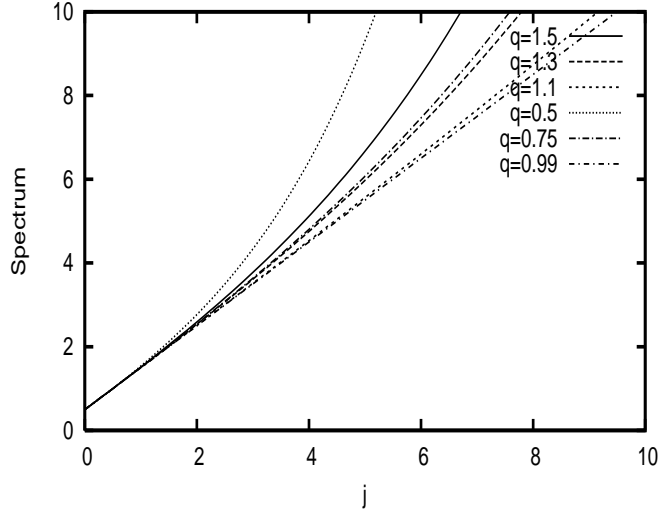}
\end{center}
\caption{Spectrum as a function of $j$ for different values of real
  $q$ and $2k=0$.}\label{graf4}
\end{figure}
\begin{figure}[htb]
\begin{center}
\includegraphics[height=7cm,width=9cm]{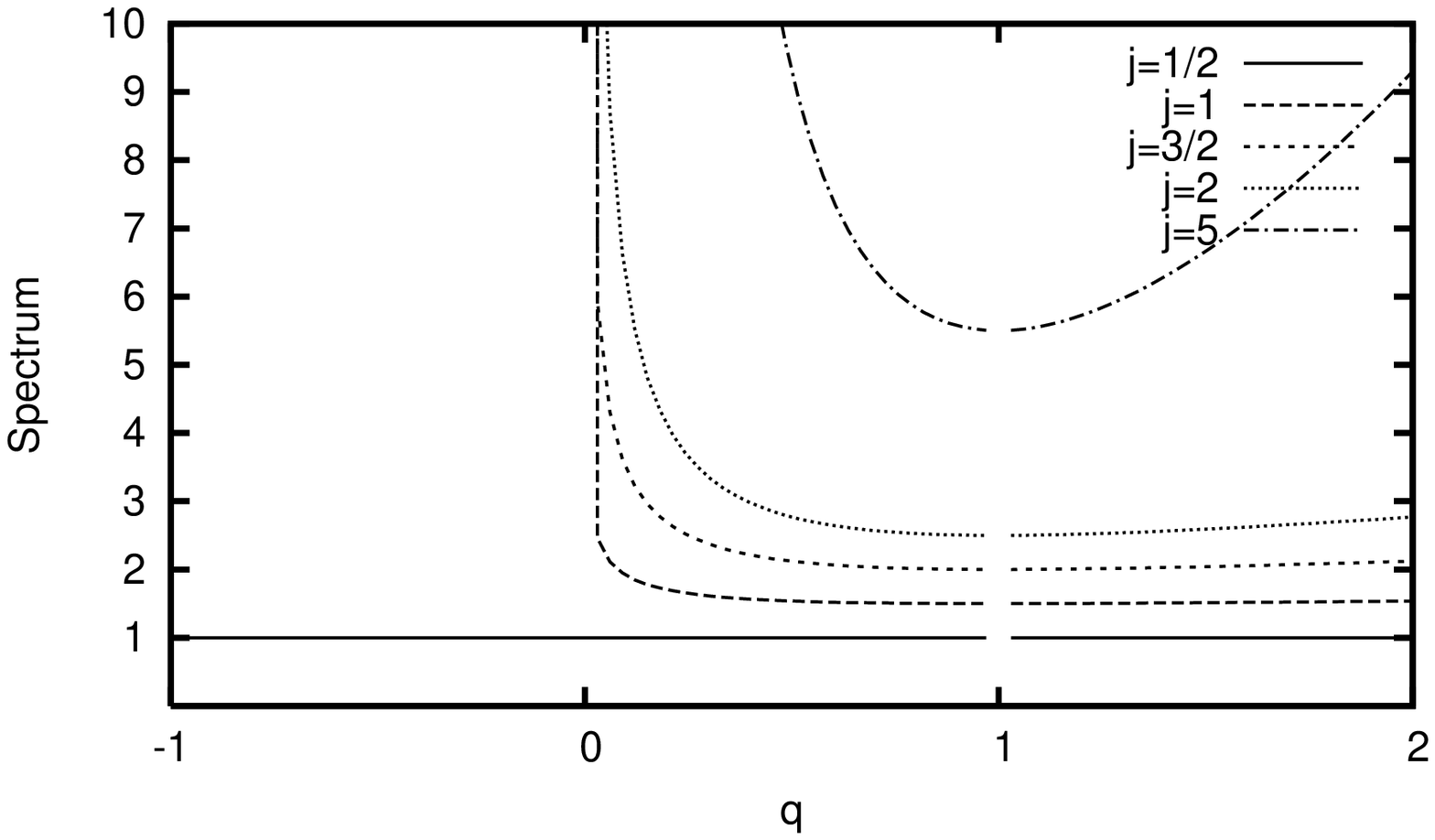}
\end{center}
\caption{Spectrum as a function of (real) $q$ for different values of
  $j$ and $2k=0$.}\label{graf5}
\end{figure}

\begin{figure}[htb]
\begin{center}
\includegraphics[height=7cm,width=9cm]{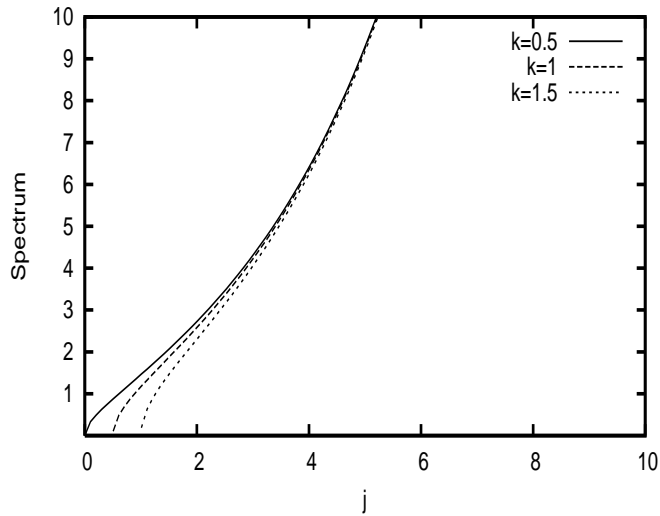}
\end{center}
\caption{Spectrum as a function of $j$ for $k\ne 0$ for real $q, q=2.$}\label{graf6}
\end{figure}
In Fig.(\ref{graf4}) we show the variation of the
spectrum with change in the spin $j$ (for fixed values of $q$). We see that the spectrum is
not doubly degenerate unlike in the previous case( where $q=e^{\fr{2\pi
    i}{p}}$). From these two plots, we see clearly that as $q\to 1$,
the spectrum approaches the fuzzy sphere result.

In Fig.(\ref{graf5}), we show the variation of the spectrum with $q$
for different values of $j$. We note that the level crossing
observed in Fig. (\ref{graf2}) for $q$ root of unity is absent for
this case of real $q$.

In Fig.(\ref{graf6}), we plot the spectrum against $j$, for given
values of $k$. We see that the spectrum becomes zero for certain values
of $j$. Since we are considering the non-zero modes, this imply
restrictions on allowed values of $k$ as in the previous case of
$q=e^{\fr{2\pi i}{p}}$.

Here too, the limits we are interested are $N\to \infty$, $q\to
1$ giving first q-deformed and then commutative spheres respectively.
As in the previous case, by sending the fuzzy cut-off to infinity, we
allow all values of spins and thus we recover the spectrum on
q-deformed sphere. Now setting the limit $q\to 1$, we get the spectrum to be
$\sqrt{(j+\half+k)(j-\half+k)}$ which is that of commutative
sphere. Whereas by keeping $N$ fixed and taking $q\to 1$ results the
spectrum corresponding to the fuzzy sphere Dirac operator.

Thus we see from the above plots that for both $q=e^{\fr{2\pi i}{p}}$
and $q=e^{2p}$, we recover the known spectrum on fuzzy as well as
commutative limits for both zero and non-zero values of topological
index.

\section{Conclusions}

In this paper, we have constructed $\uq$ invariant Dirac operator on
q-deformed fuzzy sphere and obtained its spectrum. This construction
is based on the realisation of spinor bi-modules using the creation
and annihilation operators of fermionic oscillator and a pair of
$q$-deformed bosonic operators. We have also shown that the chirality
operator constructed using the fermionic operators naturally splits
the spinor space into $\pm$ chiral subspaces. After expressing these
subspaces as the direct sum of IRRs of half-integer spins, we have
derived the operators $K_\pm$ which maps the states of one chiral
subspace to other. Using this $K_\pm$, we have constructed the Dirac
operator which maps the spinor space to itself and anti-commute with
the chirality operator as required. We obtain the eigenspinors using
the $\uq$ eigenstates belonging to the IRRs of half-integer spins.
Using the well defined action of $K_\pm$ on these states, we have
calculated the spectrum of the Dirac operator. We also obtained the
zero modes explicitly.

The Dirac operator is constructed here using the operators $K_\pm$
which maps the $\pm$ chiral subspaces to $\mp$ chiral subspaces. The
form of these operators are non-trivial compared to that in the
undeformed (i.e., $q=1$) case. Here we note that these operators in
the case of commutative sphere (see \cite{gkp}) are constructed using the
differential operators with respect to the co-ordinates of the
underlying (commutative) sphere. Thus, it is natural to expect these
derivatives to be replaced with the q-derivatives when going to the
q-sphere (which is related to $S_{qF}^2$ we considered above). It is well
known that the $q$-derivatives have highly non-trivial and asymmetric
form compared to one in the commutative
space \cite{bider,gelfand}. This fact is reflected in our operators
$K_\pm$ also. We stress the fact that the $K_\pm$ reduces to correct fuzzy
operators in the limit $q\to 1$. We also notice, using
Eqn.(\ref{keigen}), that the  q-deformed, fuzzy Laplacian \cite{amil} expressed in
terms of $K_\pm$ gives
\bea
&\fr{1}{2}(K_+ K_- + K_ - K_+)\Phi_{jkm}^j=
\frac{1}{2}([j+1+k][j-k]+[j+1-k][j+k]) \Phi_{Jkm}^j\no&\\
&= ([j][j+1]+ [k]^2 q^{\fr{2j+1}{2}} + (q^\half-q^{-\half})[k]^2(
[j]^2 q^{\fr{j+1}{2}}+[j+1]q^{\fr{j}{2}})\Phi_{Jkm}^j&.
\eea
The last two terms on the right are known as Lichnorowicz correction
to the Laplacian.
The above expression shows that in the limit $q\to 1$, the $\fr{1}{2}(K_+ K_- + K_ - K_+)$
correctly reproduces the Laplacian operator. Thus we also see that the $K_\pm$
leads to a q-deformed, fuzzy Laplacian having correct commutative limits.

We have analysed the spectrum of the Dirac operator and its various
limits. We showed that the known results are recovered in the
appropriate limits. We also showed that the spectrum of the q-deformed
fuzzy Dirac operator has many novel interesting features. We have
shown that the spectrum can be doubly degenerate depending on the
value of $p$ for the case of q being
root of unity. We have also obtained restrictions on the fuzzy
parameter and topological index coming from the deformation parameter
$q$ for the case of $q$ being root of unity. In both $q$ real as
well as root of unity, we found that the allowed values of the
topological index are constrained by the deformation parameter $q$ as
well as by fuzzy cut-off $N$.

\vspace{1cm}
\nn{\bf ACKNOWLEDGMENTS}\\
We thank A. P. Balachandran and T. R. Govindarajan for many useful
discussions and comments. ARQ and EH thank FAPESP for support through the grants 02/03247-2 and
03/09044-9 respectively. 

\nn{\bf Note added in Proof}: a different construction of Dirac operator on q-deformed sphere was carried out in \cite{kullish} and a general scheme to construct Dirac operator for coset spaces have been developed in 
\cite{balgeo}. We thank H. Steinacker for discussions and bringing the first reference to our notice.

\end{document}